\documentstyle[12pt]{article}

\begin{document}
\newcommand {\be}{\begin{equation}}
\newcommand {\ee}{\end{equation}}
\newcommand {\bea}{\begin{array}}
\newcommand {\cl}{\centerline}
\newcommand {\eea}{\end{array}}
\renewcommand {\thefootnote}{\fnsymbol{footnote}}
\baselineskip 0.65 cm
\begin{flushright}
IC/99/162\\
hep-th/9910258
\end{flushright}
\begin{center}
{\Large{\bf On the Deformation of $\Lambda$-Symmetry in 

B-field Background}}
\vskip .5cm

 M.M. Sheikh-Jabbari
\footnote{ E-mail:jabbari@ictp.trieste.it} \\

\vskip .5cm

 {\it The Abdus Salam International Center for Theoretical Physics \\

 Strada Costiera 11, Trieste, Italy}\\
\end{center}

\vskip 2cm
\begin{abstract}
In this note we will show that the $\Lambda$ symmetry, namely the $U(1)$ symmetry
of the open string sigma model which relates the B-field and the $U(1)$ gauge
field of a brane to each other, is deformed to a noncommutative version in a
constant B-field background.

\end{abstract}
\newpage
{\it Introduction}
\newline

Recently  open strings in a constant B-field background have been studied
extensively
\cite{{HD},{AAS},{HV},{NCYM},{DIPOLE},{HO1},{HO2},{SHIR},{SW},{PA},{BS}}.
The main result obtained in these papers is that the world volume of the branes
in a B-field background is a noncommutative space in terms of Connes noncommutative
geometry. 

The classical $\sigma$-model action in the B-field background, 
\be\bea{cc}
S= {1 \over 4\pi\alpha'} \int_{\Sigma} d^2\sigma \bigl[ g_{ij}
\partial_aX^{i}\partial^aX^{j}+ \epsilon^{ab} B_{ij}\partial_a
X^{i}\partial_bX^{j}+ 
{1 \over 2\pi\alpha'}\oint_{\partial \Sigma} d \tau A_i \partial_{\tau}X^i, 
\eea\ee
where $A_{\mu},$ is the $U(1)$ gauge field living on the D-brane, enjoys two $U(1)$
symmetries,

{\it $\lambda$-symmetry}:
\be
A\rightarrow A+d\lambda,
\ee
and
{\it $\Lambda$-symmetry}:
\be\left\{ \bea{cc}
B\rightarrow B+d\Lambda \\
A \rightarrow A+\Lambda
\eea\right.
\ee
It has been argued that $\lambda$-symmetry in the case of $N$ coincident D-branes is
enhanced to the $U(N)$ symmetry, 
while the $\Lambda$-symmetry is believed to remain a $U(1)$ which acts on
the $U(1)$ part of the $U(N)$ symmetry mentioned earlier \cite{BOUND}.  
In the case of the non-zero B-field background it has been argued that at the {\it
quantum level} the $\lambda$-symmetry will be deformed to a Moyal gauge symmetry
\cite{{HD},{NCYM},{SW}}, however the $\Lambda$-symmetry, has not been well
studied in these backgrounds. In all of the works mentioned
above the noncommutativity was obtained under the assumption of vanishing field
strength for the $A$
field at infinity. However, in \cite{DIR}, 
where a non-zero constant field strength for the $A$ field was considered, 
it was shown that the deformation parameter,
$\theta$, is not invariant under this classical $\Lambda$-symmetry, in other words,
one should revise this symmetry at quantum level.  

In this note, we would like to elaborate more on this symmetry.
We show that this symmetry just like the $\lambda$-symmetry, will 
be deformed to a Moyal symmetry at quantum level. It is worth noting that similar to
the arguments of \cite{SW}, this is the result of the point splitting regularization
we use.
 
{\it Calculations}

To study the $\Lambda$-symmetry at quantum level we use the Path integral
formulation and first we build the necessary tools for this calculation.

It was shown in \cite{CNY} that the two point function of open strings, $X^i(z)$
$z$ showing a point in the upper half plane, constrained to the mixed
boundary conditions, 
\be
g_{ij}(\partial-{\bar\partial})X^j+B_{ij}(\partial+{\bar\partial})X^j|_{z={\bar
z}}=0,
\ee
in the decoupling limit, i.e. $\alpha'\rightarrow 0$ and other parameter properly
scaled \cite{SW}, is
\footnote{Going to the decoupling limit makes the calculations and the
results more clean and clear.}  
\be
<X^i(z)X^j(z')>= -{1\over 2\pi}\theta^{ij}log{z-{\bar z}'\over {\bar z}-z'}+D^{ij},
\ee
where
\be
\theta^{ij}=({1\over B})^{ij},
\ee
and $D^{ij}$ is a constant which does not depend on $z, z'$ and plays no essential
role so we will set it to zero.
For the cases in which one of the operators is on the boundary, i.e. $z'={\bar
z}'=\tau'$, and the other at  $z=\tau+i\sigma,\;\;\sigma\geq 0$, the above
propagator reads as:
\be\label{KK}
<X^i(z)X^j(z')>= -{i\over \pi}\theta^{ij}\tan^{-1}{\sigma\over \tau-\tau'}\equiv
-{i\over \pi}\theta^{ij}K(\sigma,\tau;\tau').
\ee
The equation (\ref{KK}) for the values of $|\tau-\tau'|<\delta$ in the
$\delta\rightarrow 0$ limit, does not dependent on $\sigma$ and 
\be
<X^i(z)X^j(z')>= {-i}\theta^{ij}\epsilon(\tau-\tau'),
\ee
where $\epsilon(x)$ is 1 or -1 for positive or negative $x$, respectively.

Let us consider the action (1),
$$
S(B,A)=S_0+\int_{\partial\Sigma}d\tau\;A_i\partial_{\tau}X^i,
$$
where $S_0$ is the part of the action containing the string kinetic energy and
the B-field. The propagator (\ref{KK}) has been  calculated for this part.
Under the $\Lambda$-symmetry, $S(B,A)$ is transformed as
\be
\delta S=S(B+d\Lambda,A+\Lambda)-S(B,A)=
\int_{\Sigma}d\tau d\sigma\partial_{\tau}\left(\Lambda_i\partial_{\sigma}X^i\right).
\ee
As we expected it is a total time derivative, this is a classical symmetry.

In order to explore the $\Lambda$-symmetry at quantum level, we study the partition 
function of the theory under the $\Lambda$-transformations:
\be
Z(B,A)=
\int DX e^{S_0+\int_{\partial\Sigma}d\tau\;A_i\partial_{\tau}X^i},
\ee
and hence
\be\label{delta}
Z+\delta Z=\int DX e^{S_0+\int_{\partial\Sigma}d\tau\;A_i\partial_{\tau}X^i+\delta S},
\ee
Expanding (\ref{delta}) up to the first order in $A$ and $\Lambda$, we obtain
\be\bea{cc} 
\delta Z=\int DX e^{S_0}
\bigg(\int d\tau\;A_i\partial_{\tau}X^i . 
\int d\tau'
d\sigma\partial_{\tau'}\left(\Lambda_i\partial_{\sigma}X^i\right)\bigg) \\
=:\int d\tau\;A_i\partial_{\tau}X^i:\;:\int d\tau'
d\sigma\partial_{\tau'}\left(\Lambda_i\partial_{\sigma}X^i\right): \\
=\int d\tau d\sigma
d\tau'\partial_{\tau'}\bigg(:A_i(x(\tau))\Lambda_j(x(\sigma,\tau')):
:\partial_{\tau}X^i(\tau)\partial_{\sigma}X^j(\sigma,\tau'): + \\
\hspace {2.5cm} 
:A_i(x(\tau))\partial_{\sigma}X^j(\sigma,\tau'):
:\partial_{\tau}X^i(\tau)\Lambda_j(\sigma,\tau'):\bigg).
\eea\ee
The above OPE's can be evaluated by means of (\ref{KK}), however performing the 
integration in
$\tau'$, because of (8), one should note the time ordering of the operators.
In the field theory language this corresponds to  regularizing the field products by
the point splitting method \cite{SW}.
Since we are only interested in the constant $B$ and $dA$ backgrounds, for
simplicity we consider the $\Lambda$ transformations which are linear in $X^i$,
\be\label{Lam}
\Lambda_i={1\over 2}f_{ij}X_j,
\ee
where $f$ is an arbitrary constant anti-symmetric two form. Inserting (\ref{Lam})
into (12)
we find
\be
\delta Z={1\over 2}\theta^{il}\theta^{kj}f_{jl}\; \int d\tau d\sigma d\tau' 
\partial_{\tau'}\left(\partial_i A_k\; K\partial_{\sigma}\partial_{\tau}K+
\partial_k A_i\; \partial_{\sigma}K\partial_{\tau}K\right).
\ee
Integrating over sigma by parts, we obtain
\be\label{Moyal}
\delta Z={1\over 2}\theta^{il}\theta^{kj}f_{jl}\; \int d\tau d\sigma d\tau' 
\partial_{\tau'}\left(F_{ik}\; K\partial_{\sigma}\partial_{\tau}K\right),
\ee
with 
$$
F_{ik}=\partial_k A_i-\partial_i A_k.
$$   
The equation (\ref{Moyal}) can be regularized by the point splitting and then we can
perform the integration in $\tau'$. The answer  expressed in terms of the Moyal
bracket is 
\be
\delta Z=i:\int d\tau d\sigma \{A_i,\Lambda_j\}_{M.B.} \partial_{\tau}X^i(\tau) 
\partial_{\sigma}X^j(\sigma,\tau):\;\;,
\ee
where $A$ is boundary valued. 

To cancel this extra term we should deform the $\Lambda$-symmetry as the following:
\be\left\{ \bea{cc}
B\rightarrow B+d\Lambda+i\{\Lambda_i,A_j\}_{M.B.}\\
A \rightarrow A+\Lambda
\eea\right.
\ee
Although we have presented the calculations only up to  the first order in the $A$
field, along the lines of \cite{SW} it can be checked that the variations in any 
arbitrary power of $A$ will cancel out in lights of (17).

{\it Discussion}

Here we have shown that the $\Lambda$-symmetry
should be modified in the presence of the B-field,
similarly to the $\lambda$-symmetry. The interesting point 
is that the $\theta$ parameter is a function of the background $B$-field, and
therefore is changed under a $\Lambda$-transformation. 
Hence the parameter $\theta$ is not invariant under this $\Lambda$-symmetry. 

Another point we should note is that in order to use the noncommutative description
the background $A$ field should be set to zero at
infinity, by a proper $\Lambda$-transformation, i.e. $\Lambda=-A$. 
Under this $\Lambda$-transformation the B-field will transform as:
\be
B\rightarrow {\hat B}=B-dA-i\{A,A\},
\ee
recalling that $dA=F$, we find
\be\label{SW}
{\hat B}=B-(F-{1\over 4}F\theta F).
\ee
In the above relation the space-time indices have been summed over like matrix
products. The (\ref{SW}) is the same as the Seiberg-Witten map which relates the
commutative and noncommutative descriptions. In other words the results of
different regularization methods are related by this modified $\Lambda$-symmetry.
So, there is some hope that the deformed $\Lambda$-symmetry sheds light on the
Seiberg-Witten map. Studying this point and other physical consequences of this 
symmetry is postponed to future works \cite{WIP}.

{\bf Acknowledgements}

I would like to thank K. Narain for helpful discussions and D. Polyakov for reading
the manuscript.

\end{document}